\documentclass[showpacs,showkeys,11pt,
preprint,preprintnumbers,nofootinbib,
groupedaddress,superscriptaddress,amsmath,amssymb]{revtex4}
\usepackage{amsfonts}
\usepackage{amsmath}
\usepackage{graphics}
\usepackage{epsfig}
\usepackage{url}
\usepackage{multirow}
\usepackage{feynmp}
\newcommand {\be}{\begin{equation}}
\newcommand {\ee}{\end{equation}}
\newcommand {\ba}{\begin{eqnarray}}
\newcommand {\ea}{\end{eqnarray}}
\newcommand {\invfb}{$fb^{-1}$}
\newcommand {\tanb}{$\tan\beta~$}
\newcommand {\ra}{\rightarrow}

\begin{document}
\title{The t-channel Charged Higgs Production in Single Top Events at LHC}

\author{Majid Hashemi}
\email{hashemi_mj@shirazu.ac.ir}
\author{Seyyed Mohammad Zebarjad}
\author{Hossein Bakhshalizadeh}
\affiliation{Physics Department, College of Sciences, Shiraz University, Shiraz, 71946-84795, Iran}

\begin{abstract}
In this paper, the t-channel charged Higgs production at LHC is studied. Production process is a t-channel single top event with charged Higgs exchange. Therefore the signal is similar with Standard Model single top production in terms of the final state. In the first step, the signal cross section is calculated and compared to the other main production processes which are used for a heavy charged Higgs search at LHC, i.e., $pp\ra t\bar{b}H^-$ and $pp\ra H^+ \ra t\bar{b}$. In the next step, an event generation and analysis is applied on signal and background events, in order to estimate the signal significance. The signal cross section is typically smaller than the associated production ($t\bar{b}H^{-}$) and resonance production ($t\bar{b}$) by a factor of $10^{-3}$ and ranges from $10~fb$ to $1~fb$ for charged Higgs mass from 200 to 500 GeV at \tanb = 50. Due to the small cross section of signal events and large SM background, the signal significance is small even after a dedicated kinematic analysis and selection of events, however, \tanb values above 120 can be excluded at an integrated luminosity of 3000 \invfb.  
\end{abstract}
\maketitle
\section{Introduction}
While the Standard Model (SM) Higgs boson has been observed at the Large Hadron Collider (LHC) at CERN \cite{125atlas,125cms}, the question of having more than a single Higgs boson is still open. It is still not clear whether the Higgs sector is minimal and contains only a single Higgs doublet which leads to the SM Higgs boson. One of the possibilities of ``beyond the SM'' models is to have a second Higgs doublet like what is defined in Minimal Supersymmetric Standard Model (MSSM) \cite{MSSM}. The MSSM contains two CP-even neutral Higgs bosons, $h^0$, $H^0$, a CP-odd (pseudo-scalar) neutral Higgs boson, $A^0$, and pair of charged Higgs bosons, $H^\pm$. The lightest Higgs particle, $h^0$, is SM-like and is the candidate for the signal observed at LHC.\\
The recent results from LHC exclude a large parameter space of the light charged Higgs, $m_{H^\pm} < 160$ GeV, if BR$(H^{\pm}\ra \tau\nu)=1$ and heavy charged Higgs at tan$\beta>$ 50 \cite{atlasconf2,cmspas}. Such analyses are based on few production processes. The light charged Higgs is searched for through a top quark pair production with one of the top quarks decaying to charged Higgs. This assumption is valid if the charged Higgs is lighter than the top quark. The production chain is $pp\ra t\bar{t}\ra H^+bW^-\bar{b}$. Since at high \tanb, the light charged Higgs decays predominantly to a pair of $\tau\nu$, the final state of such a process contains a $\tau$ lepton which is identified using $\tau$-tagging algorithms \cite{tauidCMS}. In a previous work in \cite{jhep1}, it was shown that a $t$-channel single top production can also contribute to the light charged Higgs signal through the top quark decay to charged Higgs. \\
A charged Higgs heavier than the top quark can not be produced through $t\bar{t}$ events but is rather produced through an off-shell $t\bar{t}$ production, i.e., $pp\ra t\bar{b}H^-$. An alternative channel for heavy charged Higgs production is an $s$-channel production with subsequent decay of the charged Higgs to $t\bar{b}$ \cite{jhep2,plb}. \\
The aim of this paper is to see what can be added to the analyses of the charged Higgs at LHC by a study of the t-channel charged Higgs production. Although it turns out that the cross section of this channel is small compared to other sources of charged Higgs, the analysis is original in its own type and contains a cross section calculation as well as event generation and data analysis.\\
\section{Signal Process Cross Section}
\begin{figure}
 \begin{center}
\includegraphics[scale=0.3]{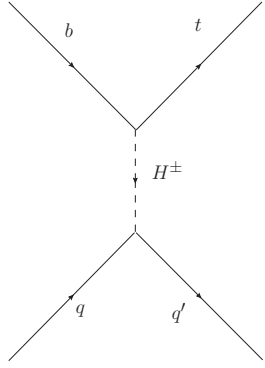}
 \end{center}
 \caption{Feynman diagram of the signal process containing a charged Higgs boson exchange. The outgoing quarks are a top quark accompanied by a light ($u,~d,~s,~c$) quark.}
 \label{tchannel}
\end{figure}

\begin{figure}
 \begin{center}
\includegraphics[scale=0.6]{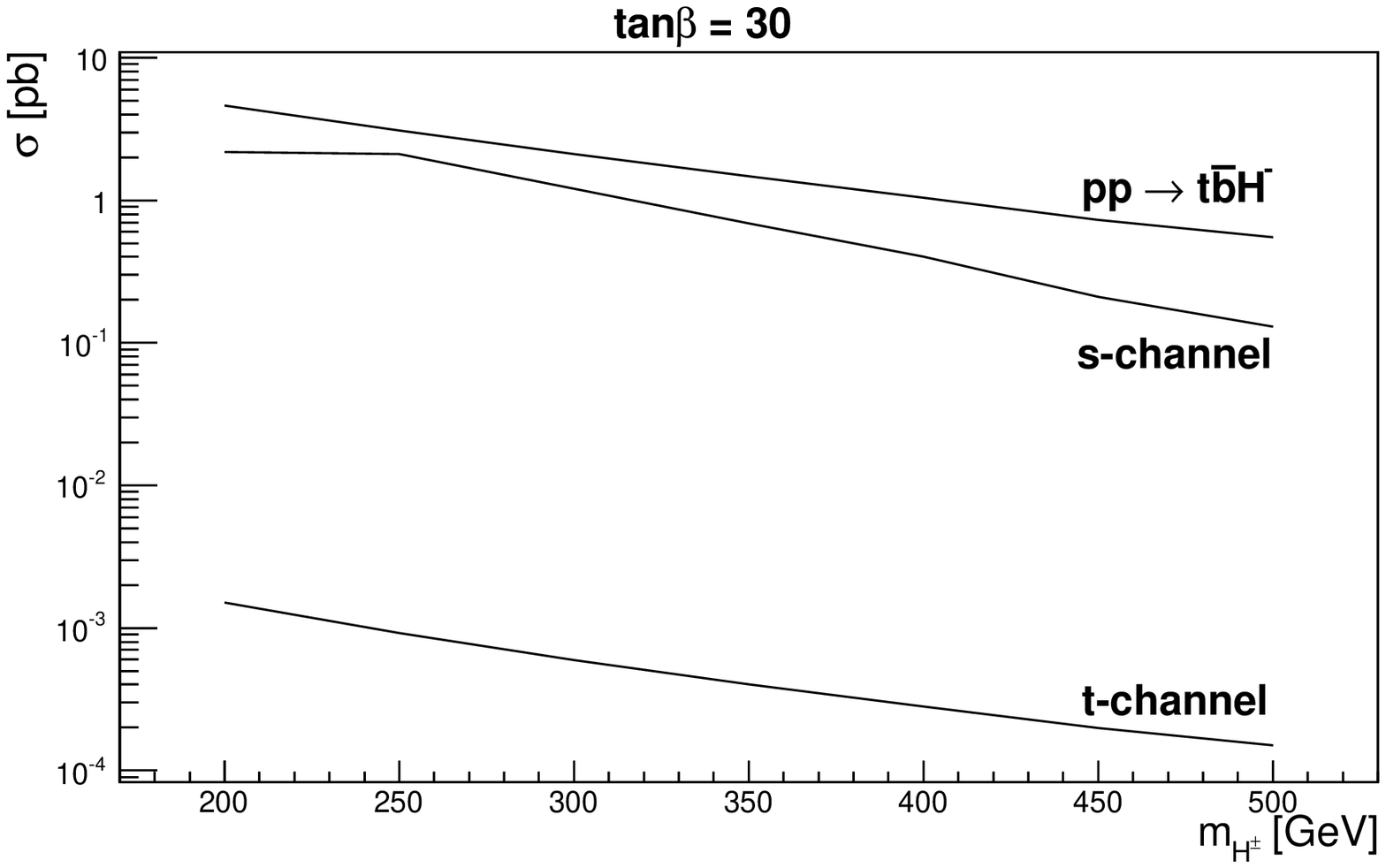}
 \end{center}
 \caption{Cross section of the $t$-channel charged Higgs exchange, the $s$-channel charged Higgs production and associated production $pp\ra t\bar{b}H^{-}$ at \tanb = 30.}
 \label{30}
\end{figure}

\begin{figure}
 \begin{center}
\includegraphics[scale=0.6]{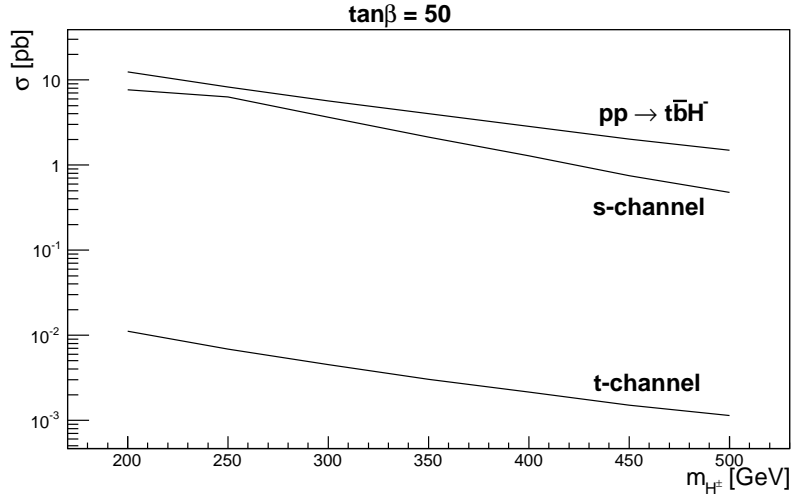}
 \end{center}
 \caption{Cross section of the $t$-channel charged Higgs exchange, the $s$-channel charged Higgs production and associated production $pp\ra t\bar{b}H^{-}$ at \tanb = 50.}
 \label{50}
\end{figure}

The signal process is a $t$-channel charged Higgs exchange with a single top quark as the final state. The top quark is accompanied by a light quark ($u,~d,~s,~c$) which produces a light jet in the detector. Figure \ref{tchannel} shows the Feynman diagram of the signal. Here we assume a Yukawa interaction Lagrangian as in Eq. \ref{lagr}, where $V_{ud}$ is the Cabbibo-Kobayashi-Maskawa (CKM) matrix element, $P_L~(P_R)$ is left (right) hand projection operator, and an implicit sum over $u$ (up type quarks) and $d$ (down type quarks) is assumed. The charged Higgs boson can be coupled to a quark pair from the same generation or from different generations. In the latter case, the off-diagonal CKM matrix element factor may suppress the coupling. Both single top and anti-top quark final states are taken into account. Therefore a total of 20 diagrams are involved. The total cross section is calculated using CompHEP 4.5.2 \cite{comphep,comphep2} with CTEQ 6.6 \cite{cteq} as the parton distribution function (PDF) of colliding protons. Results are compared with the associated production of the charged Higgs ($pp\ra t\bar{b}H^{-}$) and the $s$-channel charged Higgs production ($pp\ra H^{+}\ra t\bar{b}$) which are the main two channels for the heavy charged Higgs search at LHC. Figures \ref{30} and \ref{50} show cross sections for \tanb = 30 and 50 respectively. As is seen, the $t$-channel charged Higgs production cross section is roughly $10^{-3}$ times smaller than the other two main sources of the charged Higgs. In order to search for this signal, one has to take into account branching ratio of final state particle decays. One option would be to take the top quark decay to $W^{\pm}$ boson with a subsequent decay of $W^{\pm}$ to a lepton ($e$ or $\mu$). This choice of the signal would decrease selected events to about 20$\%$ of the initial rate. In the next section an event analysis based on kinematic peroperties of signal events is performed to evaluate the signal significance.
\be
\mathcal{L}=\sqrt{2\sqrt{2}G_F}~H^+\left[ V_{ud}(m_u \cot \beta ~\bar{u} P_L d ~+~ m_d \tan \beta ~\bar{u} P_R d ) + m_l \tan\beta ~\bar{\nu} P_R l \right] 
\label{lagr}
\ee 
\section{Event Selection and Analysis}
\begin{figure}
 \begin{center}
\includegraphics[scale=0.6]{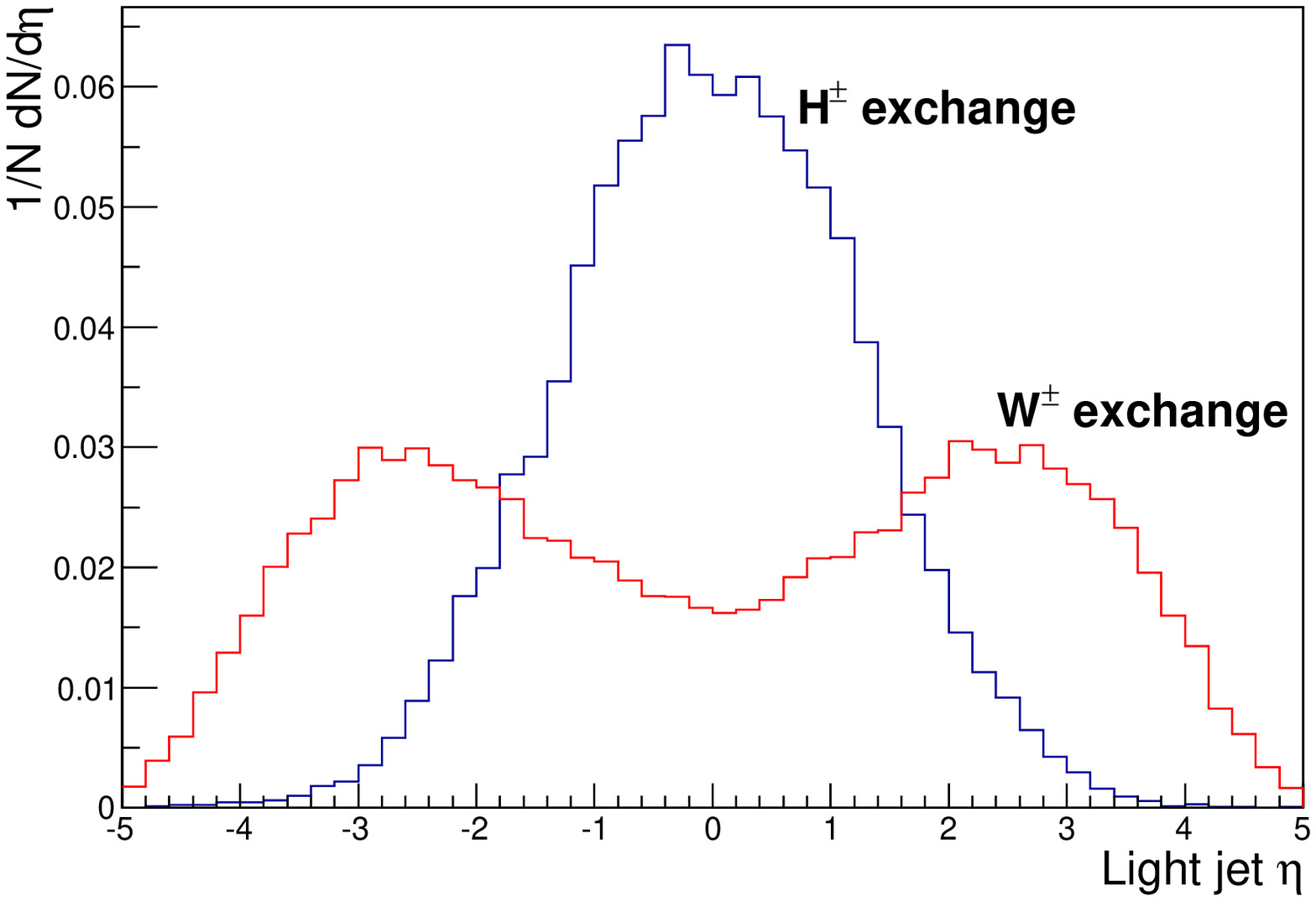}
 \end{center}
 \caption{Distribution of the light jet $\eta$ in $t$-channel single top events with $W^{\pm}$ and $H^{\pm}$ exchange. A typical charged Higgs mass of 200 GeV has been assumed.}
 \label{ljeteta}
\end{figure}

\begin{figure}
 \begin{center}
\includegraphics[scale=0.6]{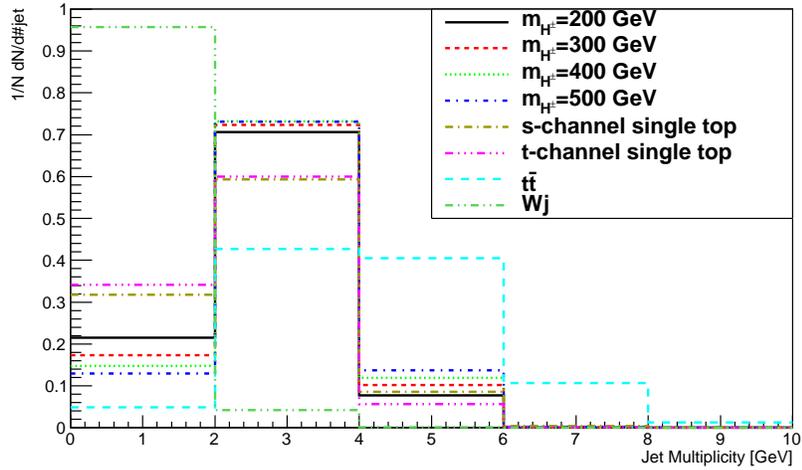}
 \end{center}
 \caption{Jet multiplicity in signal and background events.}
 \label{jetmult}
\end{figure}

\begin{figure}
 \begin{center}
\includegraphics[scale=0.6]{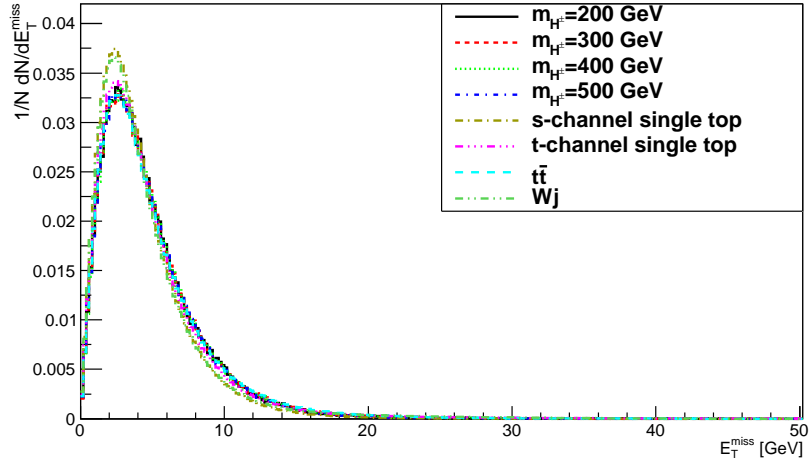}
 \end{center}
 \caption{Missing transverse energy in signal and background events.}
 \label{met}
\end{figure}

\begin{figure}
 \begin{center}
\includegraphics[scale=0.6]{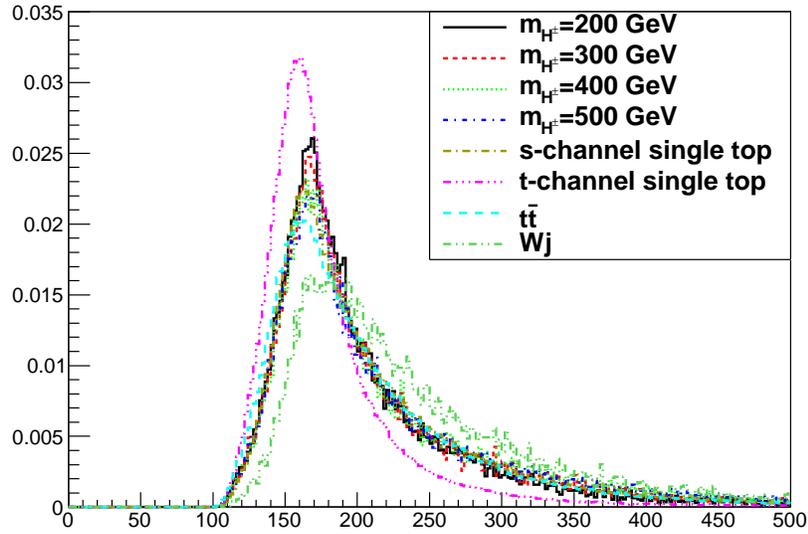}
 \end{center}
 \caption{Top quark invariant mass distribution in signal and background events.}
 \label{mtop}
\end{figure}

\begin{figure}
 \begin{center}
\includegraphics[scale=0.6]{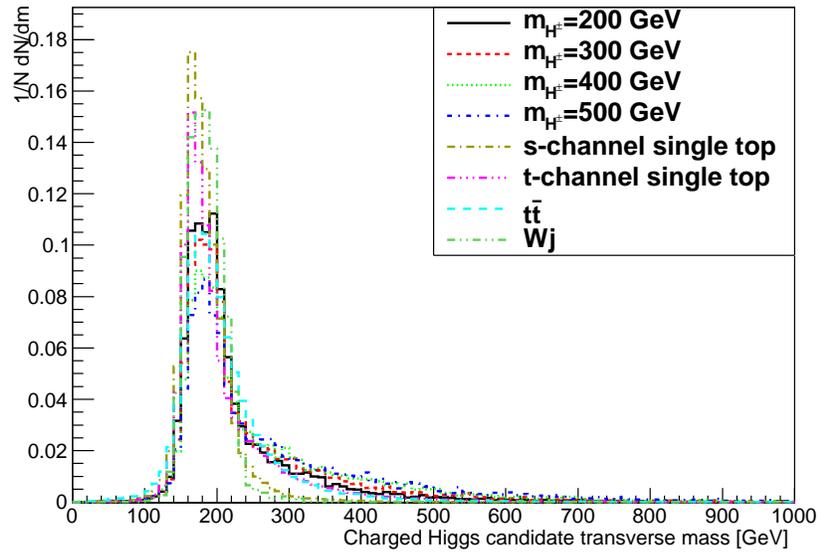}
 \end{center}
 \caption{Charged Higgs invariant mass distribution in signal and background events.}
 \label{chtm}
\end{figure}

\begin{figure}
 \begin{center}
\includegraphics[scale=0.6]{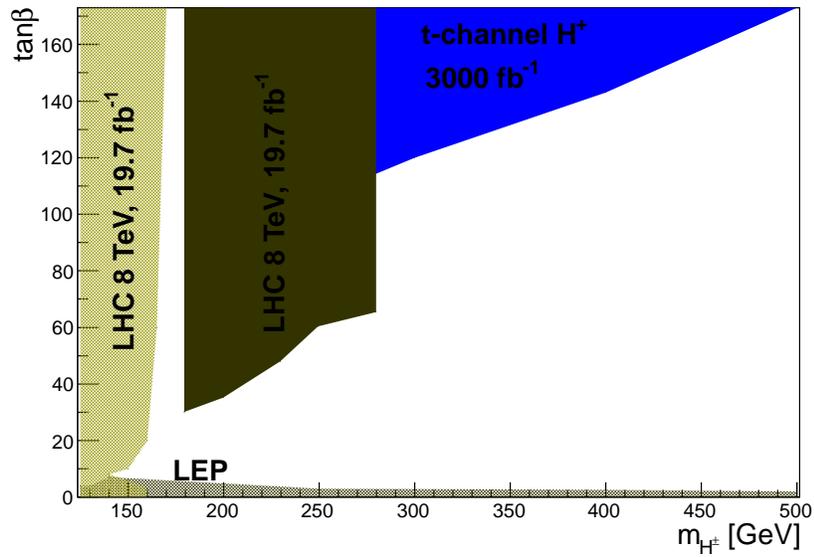}
 \end{center}
 \caption{The 95 $\%$ C.L. exclusion regions from LEP, LHC and $t$-channel charged Higgs analyses.}
 \label{2sigma}
\end{figure}

Signal events in this analysis are events with a light jet, a $b$-jet and a lepton which is taken to be an electron or muon and a neutrino which appears as missing transverse energy in the detector. Therefore the signal signature is a collection of two jets (one of them $b$-tagged) and a lepton and missing transverse energy. A similar process is the SM $t$-channel $W^{\pm}$ exchange which produces the same final state. This process is, however, different from the signal event under consideration. The $W^{\pm}$ boson is a spin-1 particle and spin considerations imply that the light quark in the final state of SM $t$-channel single top events, is emitted in forward-backward directions. This is, however, not the case for a charged Higgs exchange as the charged Higgs is spinless. Figure \ref{ljeteta} shows pseudorapidity of light jets in $t$-channel single top events with $W^{\pm}$ and $H^{\pm}$ exchange. The pseudorapidity ($\eta$) is defined as $\eta=-\ln \tan (\theta/2)$ and $\theta$ is the polar angle with respect to the beam axis. Since all other background events produce jets in the central region, which is the signal region, a cut on the light jet $\eta$ may only be helpful for SM $t$-channel single top rejection. However, other background events (single $W^{\pm}$ boson and $t\bar{t}$) are the main background and are not suppressed by such a cut. A further study of the lepton and jet transverse energies shows that signal events produce harder leptons and jets compared to single $W^{\pm}$ events but are more or less similar to other background events. Applying a cut on the jet transverse energy as jet $E_{T}>50$ GeV, results in a jet multiplicity as shown in Fig. \ref{jetmult}. As expected $t\bar{t}$ events and single $W^{\pm}$ events appear with largest and smallest number of jets per event respectively. The remaining kinematic distribution is the missing transverse energy which is shown in Fig. \ref{met}. However, no significant difference is observed between the signal and background events and therefore no cut on missing transverse energy is applied in this analysis. Before proceeding to reconstruction of signal events, an event selection is applied according to Tab. \ref{cuts}. Events which pass all these selection cuts are then used for $W^{\pm}$ boson and top quark reconstruction.\\
For the $W^{\pm}$ reconstruction, the invariant mass of the lepton and missing transverse energy is calculated. The azimuthal component of the neutrino is tuned to give the right mass of the $W^{\pm}$ boson which is set to 80 GeV. There might be no solution for the euqation which gives the W mass. In that case, the $z$-component of the neutrino momentum is set to 0. This case happens rarely, however, results in a $W^{\pm}$ mass different from the nominal value. In the next step, using $W^{\pm}$ four-momentum, and the $b$-tagged jet, top quark invariant mass is reconstructed. The $b$-tagging is a simple emulation of the algorithm which is used at LHC. It searches for $b$ or $c$ quarks near the jet, using the generator level information, and selects the jet as a $b$-jet with 60$\%$ probability if it is near a $b$ quark, or with 10$\%$ probability if close to a $c$ quark. This algorithm tries to apply the typical $b$-tagging efficiency of LHC experiments \cite{btag}. Figure \ref{mtop} shows the top quark invariant mass reconstructed using the $W^{\pm}$ and the $b$-tagged jet. An event is required to pass the top mass window which is set to the region between 150 GeV and 190 GeV. Now using four-momenta of the lepton, neutrino, $b$-jet and the light jet, the charged Higgs invariant mass is reconstructed. Since selected events have already passed hard kinematic cuts and lie in the top mass window, the charged Higgs invariant mass is obviously above the top quark mass. The top quark is fake in case of single $W^{\pm}$ events, however, due to the large cross section, a sizable amount of events pass the top quark mass window even though with a small efficiency. Figure \ref{chtm} shows the distribution of charged Higgs candidate invariant mass. As is seen, signal events tend to produce a distribution with a larger tail above 200 GeV. An event is selected if the charged Higgs candidate mass is above 200 GeV. Selection efficiencies are listed in Tabs. \ref{seleffsignal} and \ref{seleffbackground}. Using the signal and background cross sections and selection efficiencies, the signal significance can be evaluated as $N_s/\sqrt{N_b}$ where $N_s~(N_b)$ is the signal (background) number of events. Table \ref{significance} shows the signal significance for different charged Higgs masses at \tanb = 50. The signal cross section is proportional to $\tan^4\beta$ because each vertex in Fig. \ref{tchannel} acquires a \tanb factor and the matrix element squared is proportional to the fourth power of \tanb. This is a unique feature of this channel and results in a large enhancement of the signal at high \tanb. An analysis of the needed \tanb value for 95$\%$ C.L. exclusion, is performed leading to Fig. \ref{2sigma}. As is seen from Fig. \ref{2sigma}, LHC 8 TeV data, has already excluded part of the parameter space which could be excluded by this analysis. However, at high masses ($m_{H^{\pm}}>300$ GeV), very large \tanb values can be analyzed with this channel for exclusion but at a high integrated luminosity of 3000 \invfb. On the other hand, by a linear extrapolation of LHC results to high mass regions, one may expect that these accessible regions would be excluded by other analyses which require less data.

\section{Conclusions}
The $t$-channel charged Higgs exchange was studied with emphasis on the signal cross section and observability at 14 TeV LHC. The signal cross section for a heavy charged Higgs exchange in the range 200 GeV $<m_{H^{\pm}}<$ 500 GeV was calculated using CompHEP package and it turned out that, at \tanb = 50, it ranges from 10 down to 1 $fb$ in the charged Higgs mass region under study. Therefore the signal cross section is typically $10^{-3}$ times smaller than other main sources of charged Higgs which are used at LHC analyses. An event selection and analysis was performed and signal significance was obtained for different charged Higgs masses. The 95$\%$ C.L. exclusion contour shows that for a charged Higgs with $m_{H^+}=300$ GeV, \tanb above 120 can be excluded at 3000 \invfb.

\begin{figure}
 \begin{center}
\includegraphics[scale=0.6]{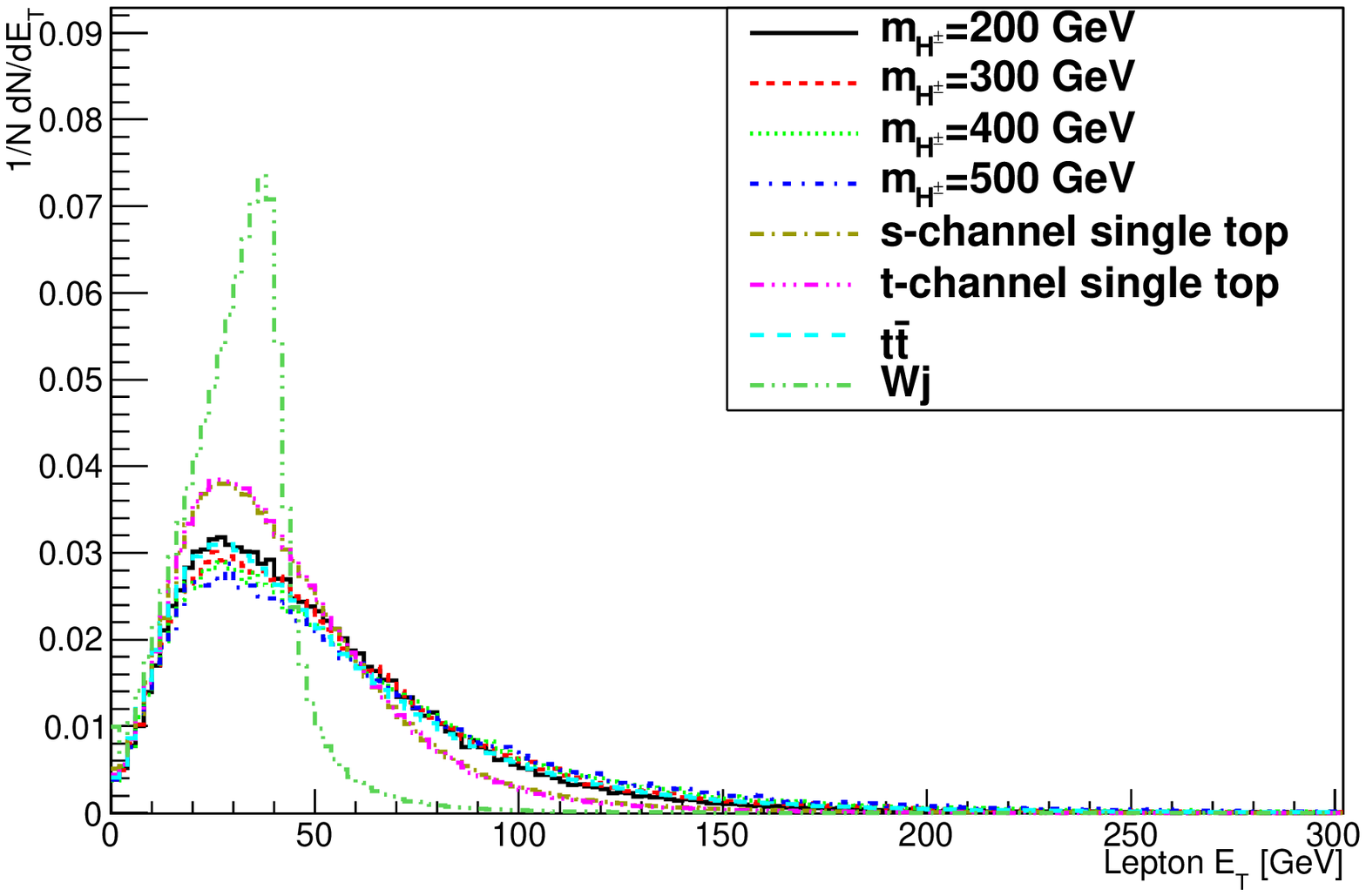}
 \end{center}
 \caption{Lepton transverse energy distribution of signal and background events.}
 \label{leptonet}
\end{figure}

\begin{figure}
 \begin{center}
\includegraphics[scale=0.6]{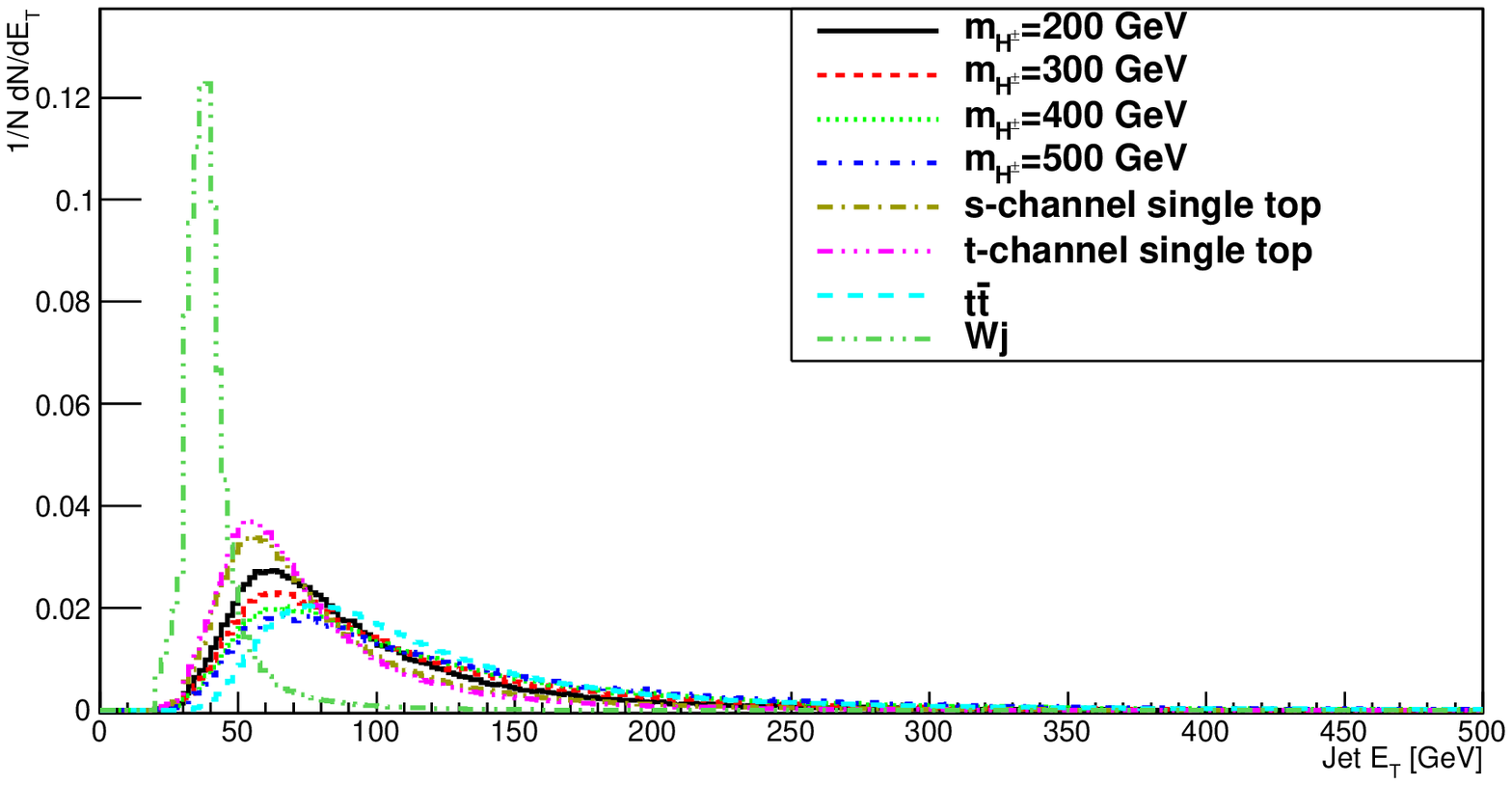}
 \end{center}
 \caption{Jet transverse energy distribution of signal and background events.}
 \label{jetet}
\end{figure}

\begin{table}
\begin{tabular}{c}
One lepton with $E_T>$ 30 GeV, $\eta<$ 2.5 \\
\hline
A total of two or three jets with $E_T>$ 50 GeV, $\eta<$ 3\\
\hline
One or two light jets $E_T>$ 50 GeV, $\eta<$ 3\\
\hline
One $b$-jet with $E_T>$ 50 GeV, $\eta<$ 3\\
\hline
\end{tabular}
\caption{Selection cuts for signal event selection. \label{cuts}}
\end{table}

\begin{table}
\begin{tabular}{|c|c|c|c|c|}
\hline
& $m_{H^{\pm}}=200$ GeV & $m_{H^{\pm}}=300$ GeV & $m_{H^{\pm}}=400$ GeV & $m_{H^{\pm}}=500$ GeV \\
\hline
$\sigma \times BR$ [fb] & 11.2 & 4.5 & 2.1 & 1.1 \\ 
\hline
\hline
One lepton & 0.63 & 0.64 & 0.65 & 0.65 \\
\hline
Two jets & 0.71 & 0.72 & 0.72 & 0.72 \\
\hline
One light jet & 0.85 & 0.84 & 0.84 & 0.83 \\
\hline
One $b$-jet & 0.62 & 0.62 & 0.62 & 0.63 \\
\hline
Top mass window & 0.42 & 0.39 & 0.37 & 0.36 \\
\hline
Charged Higgs mass cut & 1 & 0.99 & 0.99 & 0.99 \\
\hline
Total eff. & 0.099 & 0.094 & 0.09 & 0.087 \\
\hline
\hline
Events at 3000 \invfb & 3326 & 1269 & 567 & 287 \\
\hline
\end{tabular}
\caption{Signal event selection efficiencies. \label{seleffsignal}}
\end{table}
\begin{table}
\begin{tabular}{|c|c|c|c|c|}
\hline
& $t\bar{t}$ & $t$-channel single top & $s$-channel single top & W+jets \\
\hline
$\sigma \times BR$ [pb] & 690 & 197 & 9.1 & $1.7 \times 10^{5}$ \\ 
\hline
\hline
One lepton &0.62 & 0.58 & 0.52 & 0.37 \\
\hline
Two jets &0.4 & 0.6 & 0.59 & 0.062 \\
\hline
One light jet&0.8 & 0.85 & 0.87 & 0.82 \\
\hline
One $b$-jet &0.64 & 0.62 & 0.67 & 0.088 \\
\hline
Top mass window&0.35 & 0.49 & 0.38 & 0.28 \\
\hline
charged Higgs mass cut& 0.99 & 0.99 & 0.99 & 1 \\
\hline
Total eff. & 0.044 & 0.089 & 0.068 & 0.00046 \\
\hline
\hline
Events at 3000 \invfb & $91\times 10^6$ & $52\times 10^6$ & $1.8\times 10^6$ & $234\times 10^6$  \\
\hline
\end{tabular}
\caption{Background event selection efficiencies. \label{seleffbackground}}
\end{table}

\begin{table}
\begin{tabular}{|c|c|c|c|c|}
\hline
& $m_{H^{\pm}}=200$ GeV & $m_{H^{\pm}}=300$ GeV & $m_{H^{\pm}}=400$ GeV & $m_{H^{\pm}}=500$ GeV \\
\hline
Significance at 3000 \invfb & 0.17 & 0.06 & 0.03 & 0.01 \\ 
\hline
\end{tabular}
\caption{Signal significance at 3000 \invfb. \label{significance}}
\end{table}

\end{document}